\documentclass[
    ,final            % use final for the camera ready runs
  ]
  {aipproc}

\layoutstyle{6x9}

\usepackage{latexsym}
\usepackage{slashed}

\begin{document}

\title{Low-energy QCD from first principles}


\author{Marco Frasca}{
  address={Via Erasmo Gattamelata, 3 \\ 00176 Rome (Italy)},
  email={marcofrasca@mclink.it},
  homepage={http://marcofrasca.wordpress.com/}
}

%\author{<author2>}{
%  address={<common address for author2 and author3>}
%}

%\author{<author3>}{
%  address={<common address for author2 and author3>}
%  ,altaddress={<author1 address>} % additional visiting address
%}

\begin{abstract}
We show how, starting from QCD Lagrangian, a low-energy limit can be derived that is given by a non-local Nambu-Jona-Lasinio model with the form factor in close agreement with the one expected from an instanton liquid model but with all the parameters properly fixed. In the mean field approximation, the agreement with experimental data is excellent.
\end{abstract}

\maketitle

%%%%%%%%%%%%%%%%%%%%%%%%%%%%%%%%%%%%%%%%%%%%
%% MAINMATTER
%%%%%%%%%%%%%%%%%%%%%%%%%%%%%%%%%%%%%%%%%%%%

%\section{Introduction}

Knowing classical solutions to field theories can help to approach the corresponding quantum problem. For nonlinear theories these are often difficult to find and one is not able to manage a theory other than using some non-perturbative techniques as lattice computations or, in a lucky situation, with a small perturbation theory. In this paper we aim to show as a class of nonlinear field theories is amenable to analytical treatment, providing a class of classical solutions are known to be used as a starting point for perturbation theory in the infrared limit of quantum field theory. One of them is Yang-Mills theory and so, one has a technique to study this theory in the low-energy limit.

A low-energy limit for Yang-Mills theory implies that one is able to manage quantum chromodynamics (QCD) in the same limit. Indeed, if one knows the propagator of the gluonic field in this limit, the behavior of QCD at lower momenta can be obtained \cite{Frasca:2011bd, Frasca:2008zp}. The model that one recovers in such a case is a non-local Nambu-Jona-Lasinio model, as already discussed in \cite{Hell:2008cc} but with the form factor precisely fixed by QCD. This form factor agrees very well with the one obtained through an instanton liquid \cite{Schafer:1996wv} showing that, in the ground state of the theory, instantons could play a fundamental role. It is important to note that there exists another approach showing this same link between QCD and Nambu-Jona-Lasinio model \cite{Kondo:2010ts}.

In the following we show how, starting from a set of classical solutions, the low-energy limit of QCD can be recovered yielding a treatable Nambu-Jona-Lasinio model.

%\section{Classical fields}

A prototypical field theory is a massless scalar field with a simple non-linearity given by
\begin{equation}
\label{eq:cphi}
   \Box\phi+\lambda\phi^3=j.
\end{equation}
The corresponding homogeneous equation admits an exact solution $\phi = \mu\left(2/\lambda\right)^\frac{1}{4}{\rm sn}(p\cdot x+\theta,i)$
being {\rm sn} an elliptic Jacobi function and $\mu$ and $\theta$ two integration constants. In order for this solutions to hold, the following dispersion relation applies $p^2=\mu^2\sqrt{\lambda/2}$. We recognize here a free massive solution notwithstanding we started from a massless theory. We see that mass arises from the nonlinearities when $\lambda$ is taken to be finite rather than going to zero and so, standard perturbation theory just fails to recover it. Our aim is to solve the equation (\ref{eq:cphi}) in the limit $\lambda\rightarrow\infty$ and then we consider an approach devised in the '80s \cite{Cahill:1985mh} taking $\phi$ as a functional of $j$ and expanding in powers of it. The argument is consistent provided one puts
\begin{equation}
\label{eq:ps}
   \phi(x)=\mu\int d^4x'\Delta(x-x')j(x')+\mu^2\lambda\int d^4x'd^4x''\Delta(x-x')[\Delta(x'-x'')]^3j(x')+O(j(x)^3)
\end{equation}
being $\Delta(x-x')$ a solution to the nonlinear equation $\Box \Delta(x-x')+\lambda[\Delta(x-x')]^3=\mu^{-1}\delta^4(x-x')$. So, this current expansion is meaningful as a strong coupling expansion. The interesting result here is the linear term of the Green function in the current expansion. When applied to a quantum field theory this will provide a Gaussian generating functional marking a trivial theory. So, we recognize that this set of classical solutions yields, at the leading order, a trivial theory. Such a theory is mathematically manageable notwithstanding the strong nonlinearity of the theory we started from. This program can be accomplished if we know how to get the Green function and this is easily seen starting from the $d=1+0$ theory. In this case we have to solve the equation $\partial_t^2\Delta_0(t-t')+\lambda[\Delta_0(t-t')]^3=\mu^2\delta(t-t')$ and this is easily obtained \cite{Frasca:2005sx}. Then, after a Lorentz boost or a gradient expansion, the full propagator is given by
\begin{equation}
\label{eq:green}
   \Delta(p)=\sum_{n=0}^\infty(2n+1)\frac{\pi^2}{K^2(i)}\frac{(-1)^{n}e^{-(n+\frac{1}{2})\pi}}{1+e^{-(2n+1)\pi}}
   \frac{1}{p^2-m_n^2+i\epsilon}
\end{equation}
being $m_n=(2n+1)(\pi/2K(i))\left(\lambda/2\right)^{\frac{1}{4}}\mu$ and $K(i)\approx 1.3111028777$ an elliptic integral, consistently with the idea of a strong coupling expansion. Then, we would like to apply all this machinery to Yang-Mills theory but in order to show this we need to have a set of classical solutions to work with also in this case. This set of solutions would grant a trivial infrared fixed point also for this theory. Such solutions exist. This can be seen starting from the equations of motion
\begin{equation}
\scriptstyle{
\partial^\mu\partial_\mu A^a_\nu-\left(1-\frac{1}{\xi}\right)\partial_\nu(\partial^\mu A^a_\mu)+gf^{abc}A^{b\mu}(\partial_\mu A^c_\nu-\partial_\nu A^c_\mu)+gf^{abc}\partial^\mu(A^b_\mu A^c_\nu)+g^2f^{abc}f^{cde}A^{b\mu}A^d_\mu A^e_\nu = -j^a_\nu.
}
\end{equation}
and assuming again a current expansion. We note that the homogeneous equations can be solved by setting $A_\mu^a(x)=\eta_\mu^a\phi(x)$ being $\eta_\mu^a$ a set of constants. In this case we need to fix the gauge and we assume the Lorenz gauge being this equivalent to the Landau gauge in quantum field theory. So, the homogeneous equations collapse to $\partial^\mu\partial_\mu\phi+Ng^2\phi^3=-j_\phi$ and we have turned back to the previous scalar field theory (this is no more true for other gauges where the correspondence is just an asymptotic one \cite{Frasca:2009yp}). In this way, the gluon propagator in the Landau gauge is straightforwardly obtained from eq.(\ref{eq:green}) setting $\lambda=Ng^2$ and with a factor $\delta_{ab}\left(\eta_{\mu\nu}-p_\mu p_\nu/p^2\right)$.

We started with a fundamental hypothesis to support analytical developments, that is that Yang-Mills theory could admit a trivial infrared fixed point for the running coupling. Indeed, lattice computations strongly support this view as shown by the German group \cite{Bogolubsky:2009dc} from lattice at $64^4$ and $80^4$ with $\beta=5.7$ where the running coupling is seen to go to zero as momenta lower.
% (see fig. \ref{fig:rc}). 
%In fig.\ref{fig:rc} we give the result obtained by the German group
%\begin{figure}[ht!]
%\label{fig:rc}
%  \includegraphics[height=.2\textheight]{RunningCoupling}
%   \includegraphics[height=.2\textheight]{Sternbeck}
%  \caption{Running coupling for $64^4$ and $80^4$ at $\beta=5.7$ \cite{Bogolubsky:2009dc}.}
%\end{figure}
A similar result was obtained by the French group with a different definition of the infrared running coupling \cite{Boucaud:2002fx}. They show a perfect consistency with an instanton liquid model in agreement with the scenario we are depicting here.

\begin{figure}[ht!]
\label{fig:an}
  \includegraphics[height=.2\textheight]{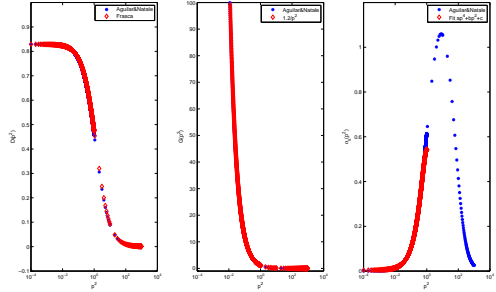}
  \caption{Comparison of our propagators with numerical solution of Dyson-Schwinger equations as in \cite{Aguilar:2004sw}. Running coupling is also shown.}
\end{figure}

Moving to quantum field theory, the generating functional for the scalar field can be managed by rescaling the space-time coordinates as $x\rightarrow\sqrt{\lambda}x$ and with a strong coupling expansion $\phi=\sum_{n=0}^\infty\lambda^{-n}\phi_n$. Then, at the leading order we will have to solve the equation $\Box\phi_0+\lambda\phi_0^3=j$ that we now know how to manage. Then, the leading order is just a Gaussian generating functional with the propagator given by eq.(\ref{eq:green}) when use is made of the approximation in eq.(\ref{eq:ps}), next-to-leading order can be also computed. We arrive at the fundamental result that the massless scalar field theory in four dimensions is infrared trivial \cite{Frasca:2010ce}. Mass spectrum is given by $m_n=(2n+1)(\pi/2K(i))\left(\lambda/2\right)^\frac{1}{4}\mu$ as expected, representing free particles with a superimposed spectrum of a harmonic oscillator. Now, turning the attention to Yang-Mills generating functional we realize that it takes the simple Gaussian form
\begin{equation}
     Z_0[j]=N\exp\left[\frac{i}{2}\int d^4x'd^4x''j^{a\mu}(x')D_{\mu\nu}^{ab}(x'-x'')j^{b\nu}(x'')\right].
\end{equation}
once we use the current expansion $A_\mu^a=\Lambda\int d^4x' D_{\mu\nu}^{ab}(x-x')j^{b\nu}(x')+O\left(1/\sqrt{N}g\right)+O(j^3)$ and the propagator $D_{\mu\nu}^{ab}(p)=\delta_{ab}\left(\eta_{\mu\nu}-\frac{p_\mu p_\nu}{p^2}\right)\Delta(p)$ being $\Delta(p)$ given by eq.(\ref{eq:green}). The spectrum in this case is that of free massive glueballs with a superimposed spectrum of a harmonic oscillator. This is consistent with our initial observation of a trivial infrared fixed point for the running coupling. Similarly, considering the ghost field, applying our approximation on the gauge field through instanton solutions, this decouples at the leading order producing a free particle propagator for a massless field. These properties of the quantum Yang-Mills field describe the so-called ``decoupling solution'' \cite{Aguilar:2004sw,Boucaud:2006if,Frasca:2007uz}. This solution is the one recovered in lattice computations \cite{Bogolubsky:2007ud,Cucchieri:2007md,Oliveira:2007px} and agrees perfectly well with the numerical solution of Dyson-Schwinger equation as given in \cite{Aguilar:2004sw} (see fig.\ref{fig:an}).

\begin{figure}[ht!]
\label{fig:ff}
  \includegraphics[height=.2\textheight]{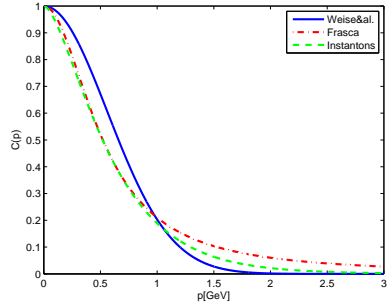}
  \caption{Comparison of our form factor with those provided in \cite{Hell:2008cc} and \cite{Schafer:1996wv} as shown in \cite{Frasca:2011bd}.}
\end{figure}

One can apply this low-energy behavior of Yang-Mills theory to the full QCD generating functional, with the propagator the one we just obtained, yielding for the action entering into it
\begin{equation}
      S=\int d^4x\left[\frac{1}{2}(\partial\sigma)^2-\frac{1}{2}m_0^2\sigma^2\right]+S_q
\end{equation}
where the $\sigma$ field arises from the gluon propagator in the Gaussian generating functional of the Yang-Mills action \cite{Frasca:2011bd} and is the contribution from the mass gap of the theory, being $m_0=(\pi/2K(i))\sqrt{\tilde\sigma}$ and $\tilde\sigma$ is the string tension ($\approx (440\ MeV)^2$). For the quark fields one has
\begin{eqnarray}
\label{eq:njl}
      S_q&=&\sum_q\int d^4x\bar q(x)\left[i{\slashed\partial}-m_q-g\sqrt{\frac{B_0}{3(N_c^2-1)}}
      \eta_\mu^a\gamma^\mu\frac{\lambda^a}{2}\sigma(x)\right]q(x) \\  
     &-&g^2\int d^4x'\Delta(x-x')\sum_q\sum_{q'}\bar q(x)\frac{\lambda^a}{2}\gamma^\mu\bar q'(x')\frac{\lambda^a}{2}\gamma_\mu q'(x')q(x)
      +O\left(\frac{1}{\sqrt{N}g}\right)+O\left(j^3\right). \nonumber
\end{eqnarray}
Now, we are able to recover the non-local Nambu-Jona-Lasinio model given in \cite{Hell:2008cc} but directly from QCD provided the form factor is
\begin{equation}
      {\cal G}(p)=-\frac{1}{2}g^2\Delta(p)=-\frac{1}{2}g^2\sum_{n=0}^\infty\frac{B_n}{p^2-(2n+1)^2(\pi/2K(i))^2\tilde\sigma+i\epsilon}
      =\frac{G}{2}{\cal C}(p)
\end{equation}
being $B_n$ obtained from eq.(\ref{eq:green}), ${\cal C}(0)=1$ and $2{\cal G}(0)=G$ the standard Nambu-Jona-Lasinio coupling, fixing in this way the value of $G$ through the gluon propagator. We compare this form factor both with the one guessed in \cite{Hell:2008cc} and the one from an instanton liquid \cite{Schafer:1996wv} in fig. \ref{fig:ff}. The result is strikingly good for the latter showing how consistently our technique represents Yang-Mills theory through instantons.

%\section{Quantum Fields}

%\section{QCD at low energies}

%\section{Numerical results}

Finally, we take the local Nambu-Jona-Lasinio limit in the non-local model and compute some observables through equations given in \cite{Klevansky:1992qe}. The agreement is absolutely excellent as seen in tab. \ref{tab:data}, notwithstanding the rough approximation, for $\Lambda=1183\ MeV$ (4d cut-off), $\tilde\sigma=(440\ MeV)^2$, $m_q=4.1\ MeV$ and $g=1.52 (\alpha_s\approx 0.18)$. Then, $G\approx 0.7854(g^2/\tilde\sigma)\approx 4.9\ GeV^{-2}$ that gives $G_{eff}\approx 2.6\ GeV^{-2}$ and so $G_{eff}\Lambda^2\approx 3.6$ being $G_{eff}=\frac{1}{m_0^2+1/G}<G$ the Nambu-Jona-Lasinio model modified by the presence of the mass gap arising after bosonization of the action (\ref{eq:njl}) as in \cite{Hell:2008cc}. We use mean field formulas as given in Klevansky \cite{Klevansky:1992qe} and refer to PDG for experimental values.
\begin{table}
\begin{tabular}{lcccc}
\hline
  \tablehead{1}{c}{b}{Observ.}
  & \tablehead{1}{c}{b}{Exp.}
  & \tablehead{1}{c}{b}{Theor.}
  & \tablehead{1}{c}{b}{Error}   \\
\hline
$m_\pi$ & 139.57018$\pm$0.00035 MeV & 139.7 MeV & 0.09\% \\
$f_\pi$ & 130.41$\pm$0.03$\pm$0.20 MeV & 131.5 MeV & 0.9\% \\
$-\langle\bar u u\rangle^\frac{1}{3}=-\langle\bar d d\rangle^\frac{1}{3}$ & 230$\pm$10  MeV (sum rules) & 274 MeV & 16\% \\
$g_{\pi qq}$ & - & 2.3 & - \\
\hline
\end{tabular}
\caption{Comparison with experimental values in the contact approximation \cite{Klevansky:1992qe}}
\label{tab:data}
\end{table}

%\section{Conclusions}

We have shown how, starting from QCD Lagrangian, it is possible to derive a non-local Nambu-Jona-Lasinio model with all parameters properly fixed. We were able to compute some meaningful observables in close agreement with experimental data even if in a mean field approximation.

%%%%%%%%%%%%%%%%%%%%%%%%%%%%%%%%%%%%%%%%%%%%%%%%
%% BACKMATTER
%%%%%%%%%%%%%%%%%%%%%%%%%%%%%%%%%%%%%%%%%%%%%%%%

%\begin{theacknowledgments}
I would like to thank Marco Ruggieri for very enlightening comments and the code for numerical Dyson-Schwinger equations.
%\end{theacknowledgments}



\bibliographystyle{aipproc}   % if natbib is available

%\bibliography{FrascaProc}

%  \typeout{******************************************}
%  \typeout{** Please run "bibtex \jobname" to optain}
%  \typeout{** the bibliography and then re-run LaTeX}
%  \typeout{** twice to fix the references!}
%  \typeout{******************************************}

%\end{document}

%%%%%%%%%%%%%%%%%%%%%%%%%%%%%%%%%%%%%%%%%%%
%% The following lines show an example how to produce a bibliography
%% without the help of the BibTeX program. This could be used instead
%% of the above.
%%%%%%%%%%%%%%%%%%%%%%%%%%%%%%%%%%%%%%%%%%%

\end{document}